\documentclass{PoS}
\usepackage{slashed,amsmath,amssymb}

\title{Mass anomalous dimension from large N twisted volume reduction}

\ShortTitle{Mass anomalous dimension from large N twisted volume reduction}

\author{Margarita Garc\'{\i}a P\'erez\\
        Instituto de F\'{\i}sica Te\'orica UAM/CSIC, Universidad Aut\'onoma de Madrid, 
E-28048-Madrid, Spain\\
        E-mail: \email{margarita.garcia@uam.es}}

\author{Antonio Gonz\'alez-Arroyo\\
Instituto de F\'{\i}sica Te\'orica UAM/CSIC and  Departamento de F\'{\i}sica Te\'orica, C-15,
Universidad Aut\'onoma de Madrid, E-28049-Madrid, Spain \\
        E-mail: \email{antonio.gonzalez-arroyo@uam.es}}

\author{\speaker{Liam Keegan} \\%
        Instituto de F\'{\i}sica Te\'orica UAM/CSIC, Universidad Aut\'onoma de Madrid, 
E-28048-Madrid, Spain\\
        E-mail: \email{liam.keegan@uam.es}}

\author{Masanori Okawa\\
Graduate School of Science, Hiroshima University, 
Higashi-Hiroshima, Hiroshima 739-8526, Japan\\
        E-mail: \email{okawa@sci.hiroshima-u.ac.jp}}

\abstract{In this work we consider the SU(N) gauge theory with two Dirac fermions in the adjoint representation, in the limit of large N. Taking advantage of large N twisted volume reduction we do this on a single site lattice, but we should still get infinite-volume physics in the large N limit. We describe our progress in extracting the mass anomalous dimension from the eigenvalue distribution of the adjoint Dirac operator, using data for N up to 289.}

\FullConference{31st International Symposium on Lattice Field Theory LATTICE 2013\\
                 July 29 - August 3, 2013\\
                 Mainz, Germany}

\begin{document}

\section{Introduction}
The SU(2) gauge theory with two adjoint Dirac fermions, known as Minimal Walking Technicolor (MWT)~\cite{Sannino:2004qp,Luty:2004ye}, has been the subject of many lattice studies, all of which have found it to be a conformal theory with a fairly small mass anomalous dimension~\cite{Bursa:2009we,DelDebbio:2010hx,DeGrand:2011qd,Catterall:2011zf}. The most recent measurement obtained by fitting the mode number of the Dirac operator gave a very precise value~\cite{Patella:2012da}. The mode number method has also been used to follow the running of $\gamma$ over a range of energy scales for the SU(3) theory with many light fundamental fermions~\cite{Cheng:2013eu}.

The large N version of MWT, the SU(N) gauge theory with two adjoint fermions, is interesting for several reasons. From a phenomenological point of view, it is expected to be similar to the SU(2) theory, for example the first two universal orders of perturbation theory predict an infrared fixed point with a mass anomalous dimension that is independent of $N$. From a numerical point of view, simulating the theory on a single site lattice using large N twisted volume reduction~\cite{GonzalezArroyo:1982hz,GonzalezArroyo:2010ss} makes it feasible to simulate at values of N that would be prohibitively expensive on a conventional $L^4$ lattice. We use the action described in Refs.~\cite{Gonzalez-Arroyo:2013bta,Gonzalez-Arroyo:2013proc} to simulate a single site lattice at N up to 289, which is equivalent to a lattice of size $L^4=17^4$. We then extract a measurement of the mass anomalous dimension from a fit to the mode number of the Dirac operator.

\section{Method}
In a mass--deformed conformal field theory (mCFT), the spectral density $\rho$ of the Dirac operator at small eigenvalues $\omega$ scales as~\cite{DelDebbio:2010ze} 
\begin{equation}
\label{eq:rho}
\lim_{m\rightarrow 0}\lim_{V\rightarrow\infty}\rho(\omega) \propto \omega^{\frac{3-\gamma_*}{1+\gamma_*}},
\end{equation}
where $\gamma_*$ is the mass anomalous dimension at the infrared fixed point (IRFP), $V$ is the lattice volume, and $m$ is the mass.

On the lattice we measure the eigenvalues $\Omega^2$ of the massive hermitian Dirac operator $M=m^2 - \slashed{D}^2$, which are related to $\omega$ by $\omega = \sqrt{\Omega^2 - m^2}$. The mode number $\overline{\nu}(\Omega)$ is simply the number of eigenvalues of this operator below some value $\Omega^2$ divided by the volume (the volume being $N^2$ in this case), and can be written as
\begin{equation}
\overline{\nu}(\Omega) = 2 \int_0^{\sqrt{\Omega_{IR}^2-m^2}}\rho(\omega)\,\, d\omega + 2\int_{\sqrt{\Omega_{IR}^2-m^2}}^{\sqrt{\Omega^2-m^2}}\rho(\omega)\,\, d\omega,
\end{equation}
where the integral has been split into two parts. The first part will be affected by finite volume and/or finite mass effects, but we can insert Eq.~\ref{eq:rho} in the second term. Integrating and writing in lattice units gives the fit function
\begin{equation}
\label{eq:fitIII}
a^{-4}\overline{\nu}(\Omega) \simeq a^{-4}\overline{\nu}(\Omega_{IR})-A\left[(a\Omega_{IR})^2-(am)^2\right]^{\frac{2}{1+\gamma_*}} + A\left[(a\Omega)^2-(am)^2\right]^{\frac{2}{1+\gamma_*}}.
\end{equation}
From this one can extract $\gamma_*$ by fitting the three free parameters of this fit, $A,am,\gamma_*$, in some intermediate range of eigenvalues $\Omega_{IR}<\Omega<\Omega_{UV}$, whilst maintaining the separation of scales on the lattice,
\begin{equation}
\frac{1}{a\sqrt{\mathrm{N}}} \ll m \ll \Omega_{IR} < \Omega < \Omega_{UV} \ll \frac{1}{a}.
\label{eq:scales}
\end{equation}

\section{Simulation Details}
Full details of the action and the implementation are given in Refs.~\cite{Gonzalez-Arroyo:2013bta,Gonzalez-Arroyo:2013proc}. Here we use 20-40 configurations for $b=0.35,0.36$ and $N=121,289$, each separated by 125 molecular dynamics updates. We measure the lowest 1000 eigenvalues on the N=121 configurations, and the lowest 2000 eigenvalues on the N=289 configurations. For $b=0.36$ we also do simulations with higher statistics for $N=16,25,49$, to investigate the $1/N$ corrections. We cannot do the same for $b=0.35$ because there is a strong coupling phase at small $b$, and the simulations at $b=0.35$ for smaller $N$ fall into this wrong phase.

Physical quantities in the twisted reduced model depend on $b$, $\kappa$, $N$ and $k$. In the large $N$ limit, reduction implies that the results should be independent of $k$ provided the center symmetry is not broken. According to Ref.~\cite{GonzalezArroyo:2010ss}, this demands that the limit has to be taken keeping $k/\sqrt{N}>1/9$ and $\tilde{\theta}=\bar{k}/\sqrt{N}>\tilde{\theta}_c$. The integer $\bar{k}$ is defined by the relation $\bar{k}k=1 \bmod \sqrt{N}$. The $1/N$ corrections are expected to depend slightly on $\tilde{\theta}$, so that extrapolations are better done keeping $\tilde{\theta}$ as constant as possible, which is not always easy to do for small $N$. Tab.~\ref{tab:k} shows the values 
of $N$, $k$ and $\tilde{\theta}$ used in this work. 

\begin{table}
\begin{center}
\begin{tabular}{c|c|c|c|c|c}
& $N=16$ & $N=25$ & $N=49$ & $N=121$ & $N=289$ \\
\hline
$k$ & 1 & 2 & 3 & 3 & 5 \\
$\tilde\theta$ & 0.25 & 0.40 & 0.29 & 0.36 & 0.41 \\
\end{tabular} 
\end{center}
\caption{Value of the flux $k$ used for each $N$, along with the corresponding value of $\tilde\theta = \bar k / \sqrt{N}$}
\label{tab:k}
\end{table}

\section{Results}
\subsection{Reduction}
\begin{figure}
  \centering
    \includegraphics[angle=270,width=7.0cm]{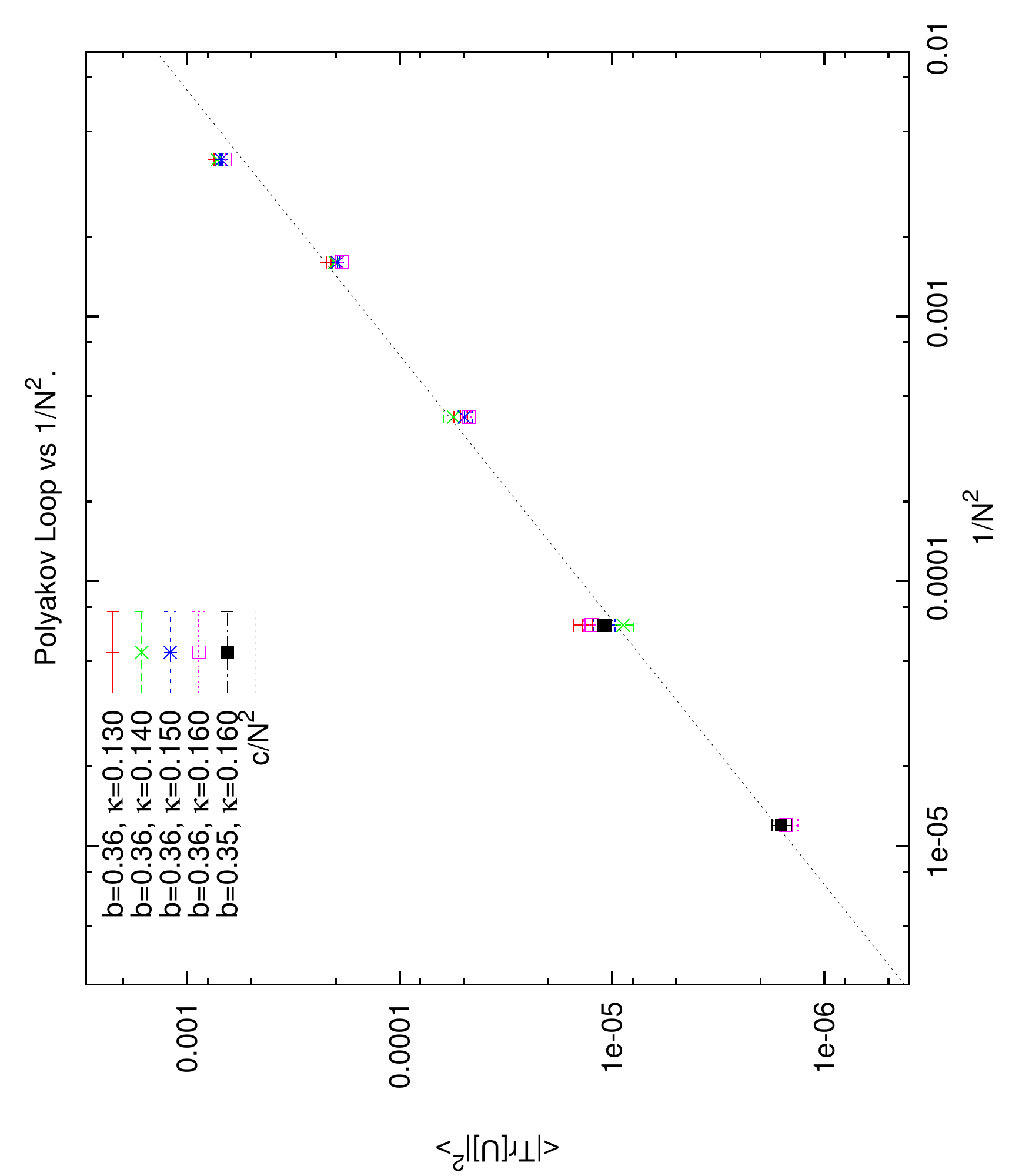}
  \caption{Polyakov loop $\left|Tr[U]\right|^2$ vs $1/N^2$, showing that center symmetry is preserved, and thus that reduction holds, in the large N limit.}
  \label{fig:poly}
\end{figure}

For reduction to hold center symmetry must be preserved, which means that the modulus of the Polyakov loop must go to zero in the $N\rightarrow\infty$ limit.  Fig.~\ref{fig:poly} shows the modulus squared of the Polyakov loop vs $1/N^2$, confirming that reduction holds in the large N limit.

% \begin{figure}
%   \centering
%     \includegraphics[angle=270,width=10.0cm]{figs/plaq.pdf}
%   \caption{Plaquette vs $1/\sqrt{N}$. $1/N$ corrections become larger for lighter masses (larger values of $\kappa$). }
%   \label{fig:plaq}
% \end{figure}

\subsection{Lowest Eigenvalue}

\begin{figure}
  \centering
    \includegraphics[angle=270,width=7.0cm]{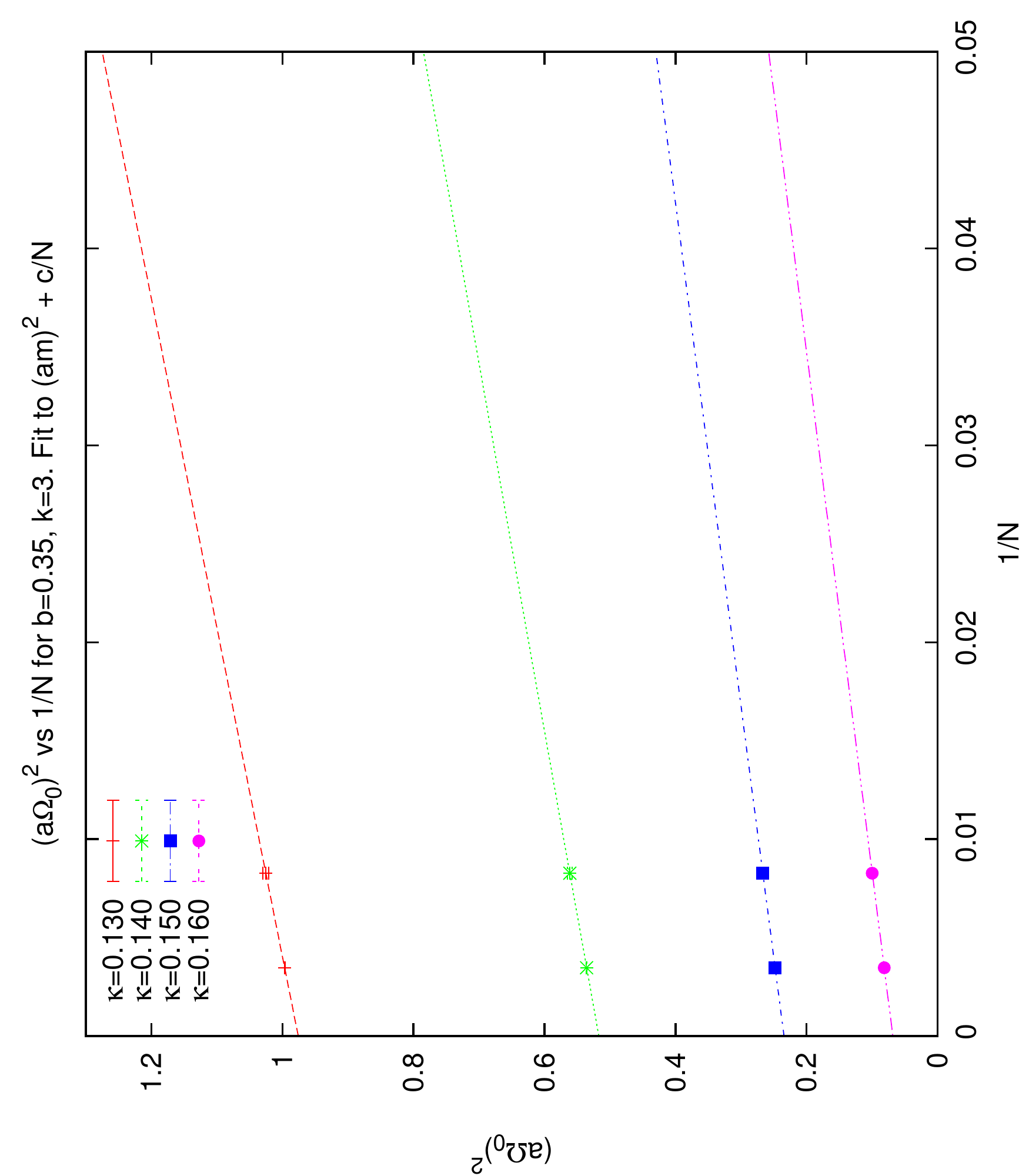}\includegraphics[angle=270,width=7.0cm]{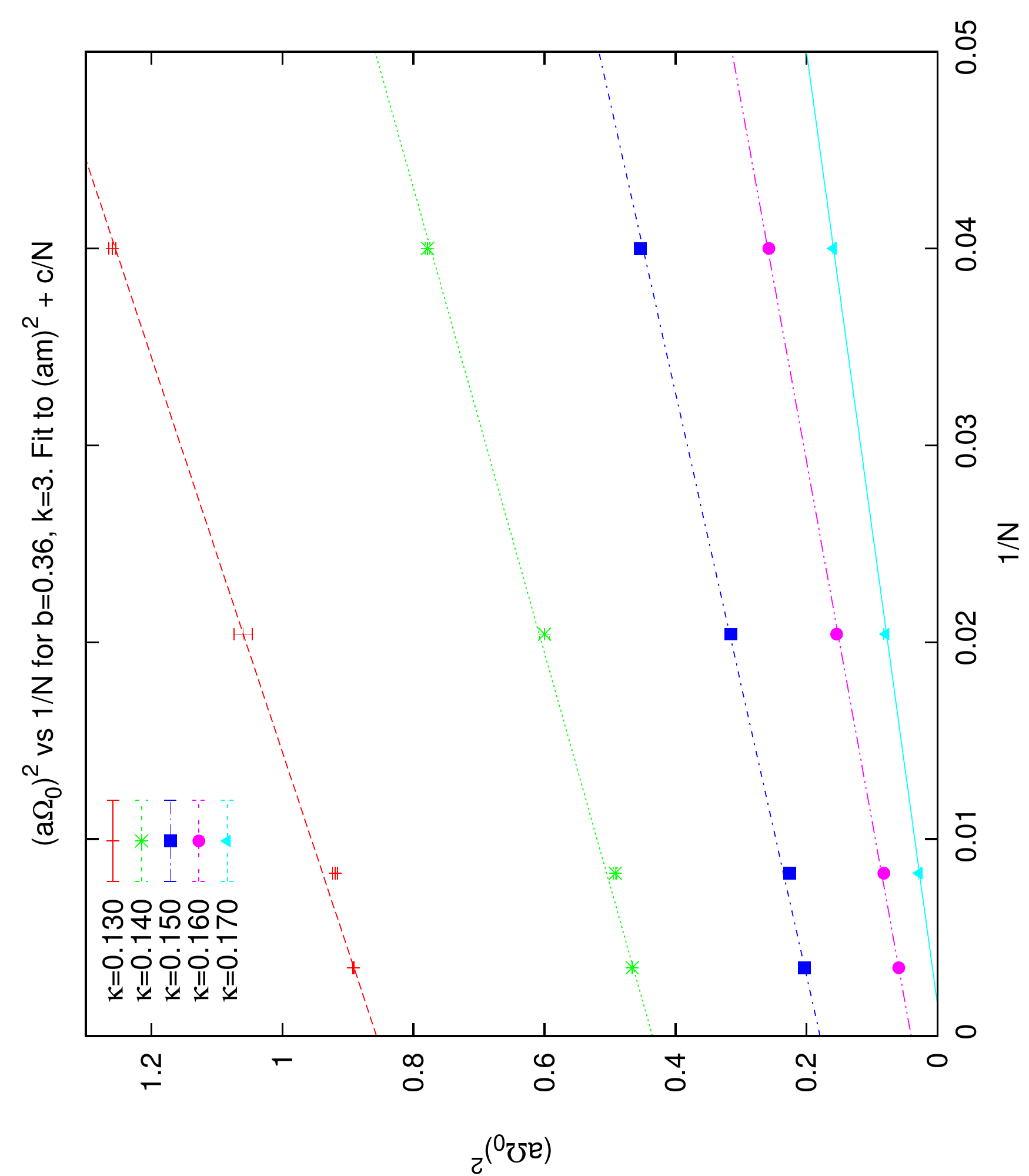}
  \caption{Lowest eigenvalue squared for $b=0.35$ (left), $b=0.36$ (right), $(a\Omega_0)^2$, vs $1/N$. The fitting form used is $(a\Omega_0)^2 = (am)^2 + c/N$.}
  \label{fig:msqvsN}
\end{figure}

In twisted reduction the boundary conditions prohibit a zero momentum state, and for a trivial background gauge field and to leading order in $1/N$, the lowest eigenvalue scales as
\begin{equation}
(a\Omega_0)^2 = (am)^2 + (ap)^2,
\end{equation}
where $ap = 2\pi/L = 2\pi/\sqrt{N}$. It turns out this form also fits the non--perturbative data fairly well, with the substitution $(ap)^2=c/N$ where $c$ is a free fit parameter, as shown in Fig.~\ref{fig:msqvsN}. In the infinite--N limit the lowest eigenvalue corresponds to the parameter $am$ in Eq.~\ref{eq:fitIII}. The values for $am$ determined from Fig.~\ref{fig:msqvsN} in this way are shown in Tab.~\ref{tab:am}.

\begin{table}
\begin{center}
\begin{tabular}{c|c|c}
b & $\kappa$ & $(am)^2$\\
\hline
0.35 & 0.160 & 0.070(1)\\
0.36 & 0.160 & 0.040(1)
\end{tabular} 
\end{center}
\caption{$(am)^2$ as determined from an extrapolation of the lowest eigenvalue in $1/N$. The quoted error is statistical only. There is an additional systematic error due to $\tilde{\theta}$ varying slightly with $N$.}
\label{tab:am}
\end{table}

\subsection{Eigenvalue Spectrum}
\begin{figure}
  \centering
    \includegraphics[angle=270,width=7cm]{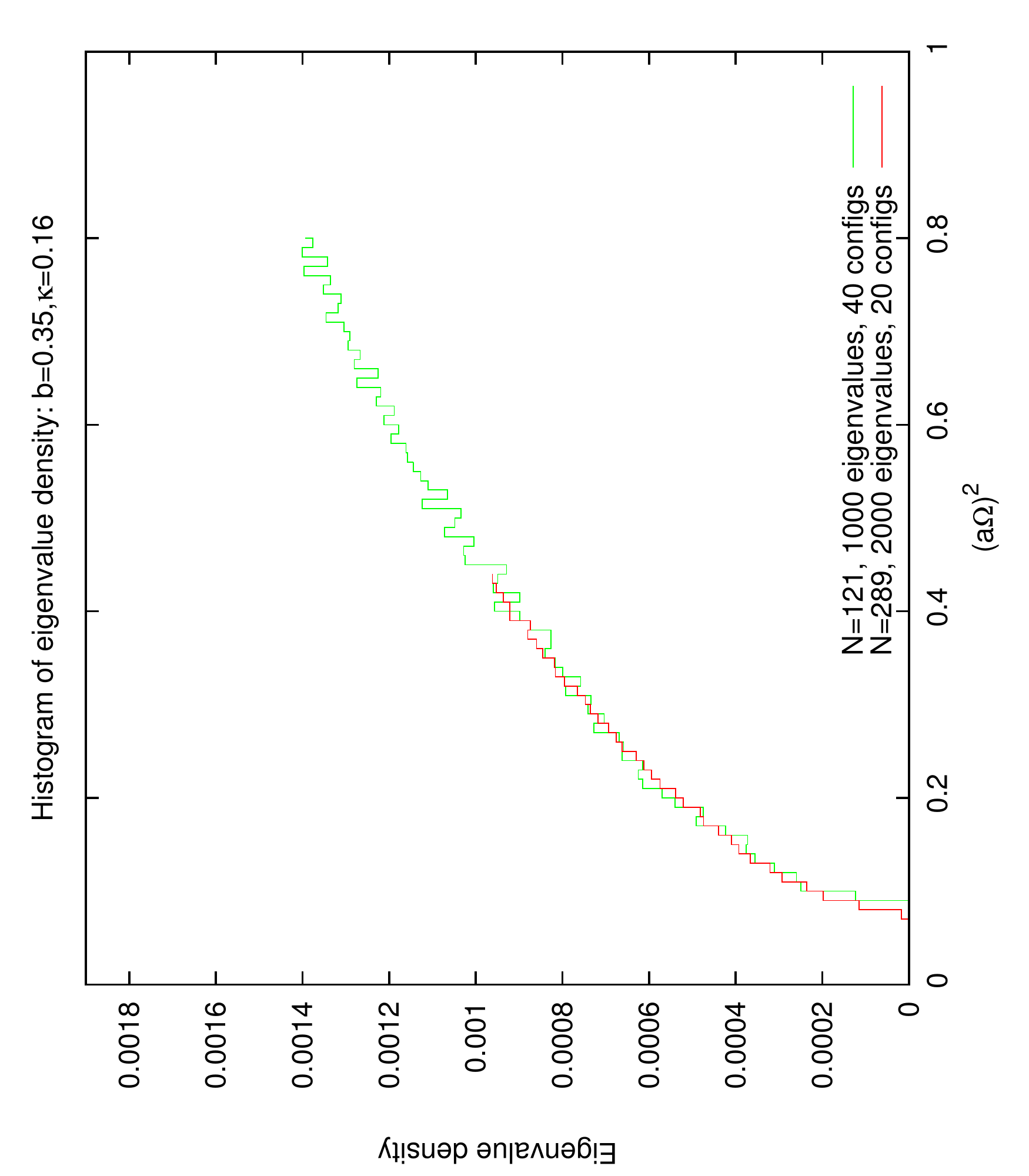}\includegraphics[angle=270,width=7cm]{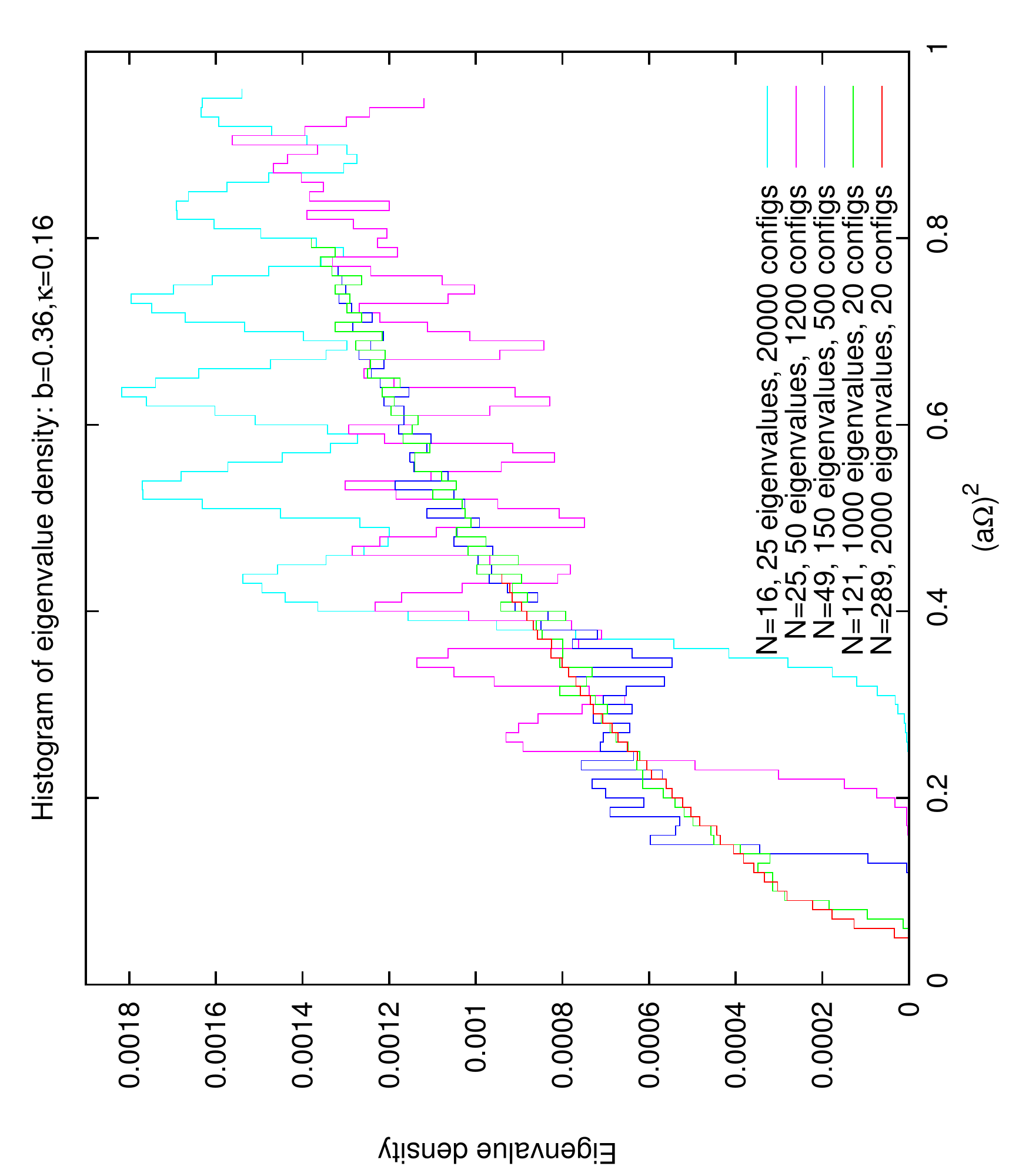}
  \caption{Eigenvalue spectrum histogram for $b=0.35$ (left), $b=0.36$ (right), $\kappa=0.16$, for various $N$. For the smaller values of $N$ there is a region of small eigenvalues that is clearly dominated by finite--N effects, but as N is increased this region is pushed to lower eigenvalues, and above this region there is agreement between different values of N.}
  \label{fig:histogram}
\end{figure}

Fig.~\ref{fig:histogram} shows a histogram of the number of eigenvalues as a function of $(a\Omega)^2$, for $b=0.35,0.36$, $\kappa=0.160$. Comparing different values of N, we see agreement between N=289 and N=121 in all but the first two bins. For N=49, for $(a\Omega)^2 \lesssim 0.4$ the behaviour is qualitatively different, but for higher eigenvalues we again see agreement with larger values of N. For N=25 and N=16 there is no agreement, except possibly at the largest eigenvalues shown, and there is also a clear oscillation, or clumping of the eigenvalues, which is a finite N effect.

The main point is that, even if low eigenvalues are affected by finite volume effects, for sufficiently large eigenvalues the eigenvalue density should agree between different N. The mode number, which is the integral of this density, will differ by some constant term due to the difference in the contribution from the small eigenvalues. However, a fit to Eq.~(\ref{eq:fitIII}) in the region where the densities agree should still give consistent values for $A$, $am$ and $\gamma$. Based on this observation we choose to use $\kappa=0.160$ for the mode number fits, as it is the lightest mass for which we have data at $N=289$, and after excluding the lowest eigenvalues finite volume effects for the mode number are under control.

\subsection{Mode number fit}

\begin{figure}
  \centering
    \includegraphics[angle=270,width=7.0cm]{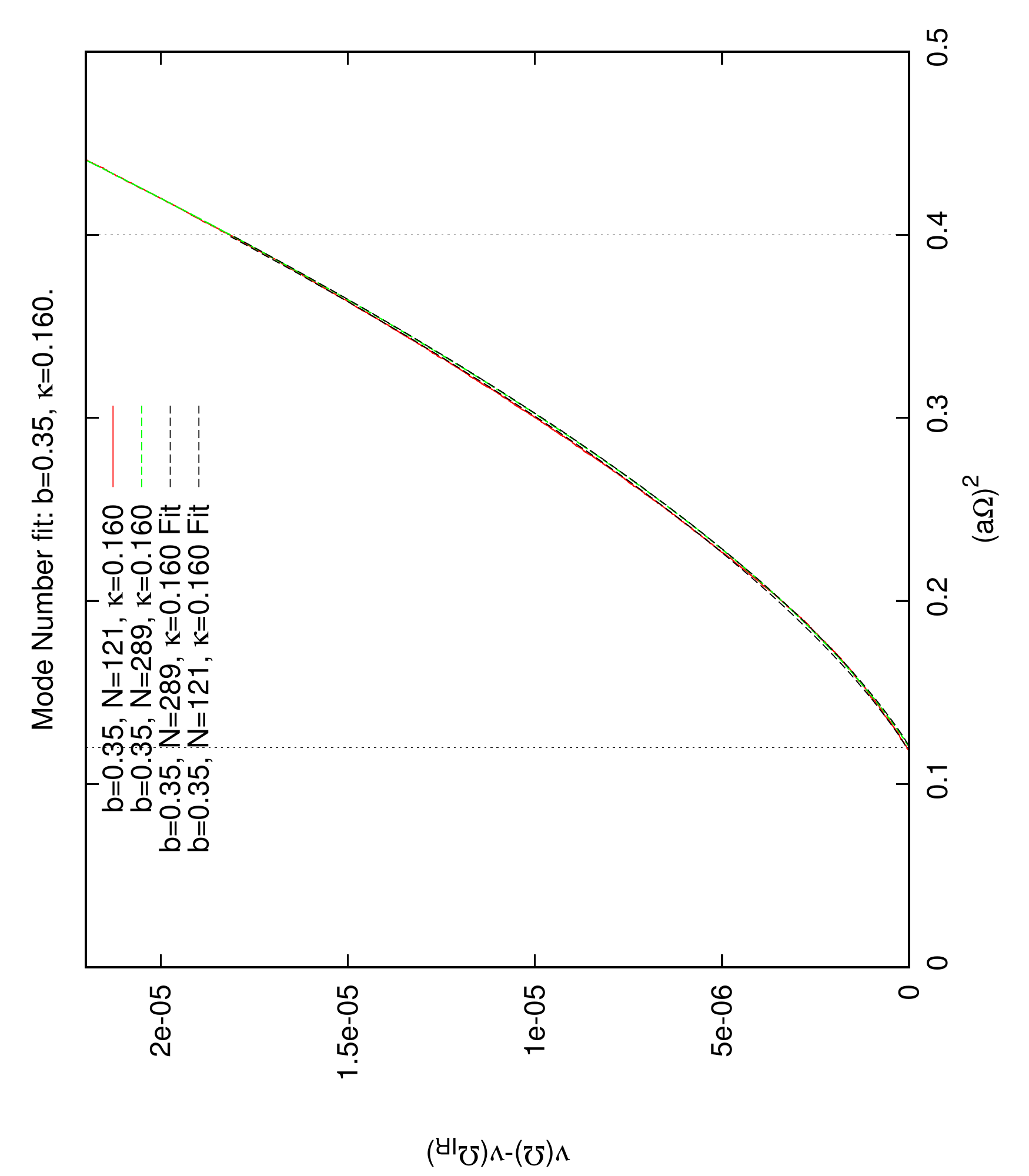}\includegraphics[angle=270,width=7.0cm]{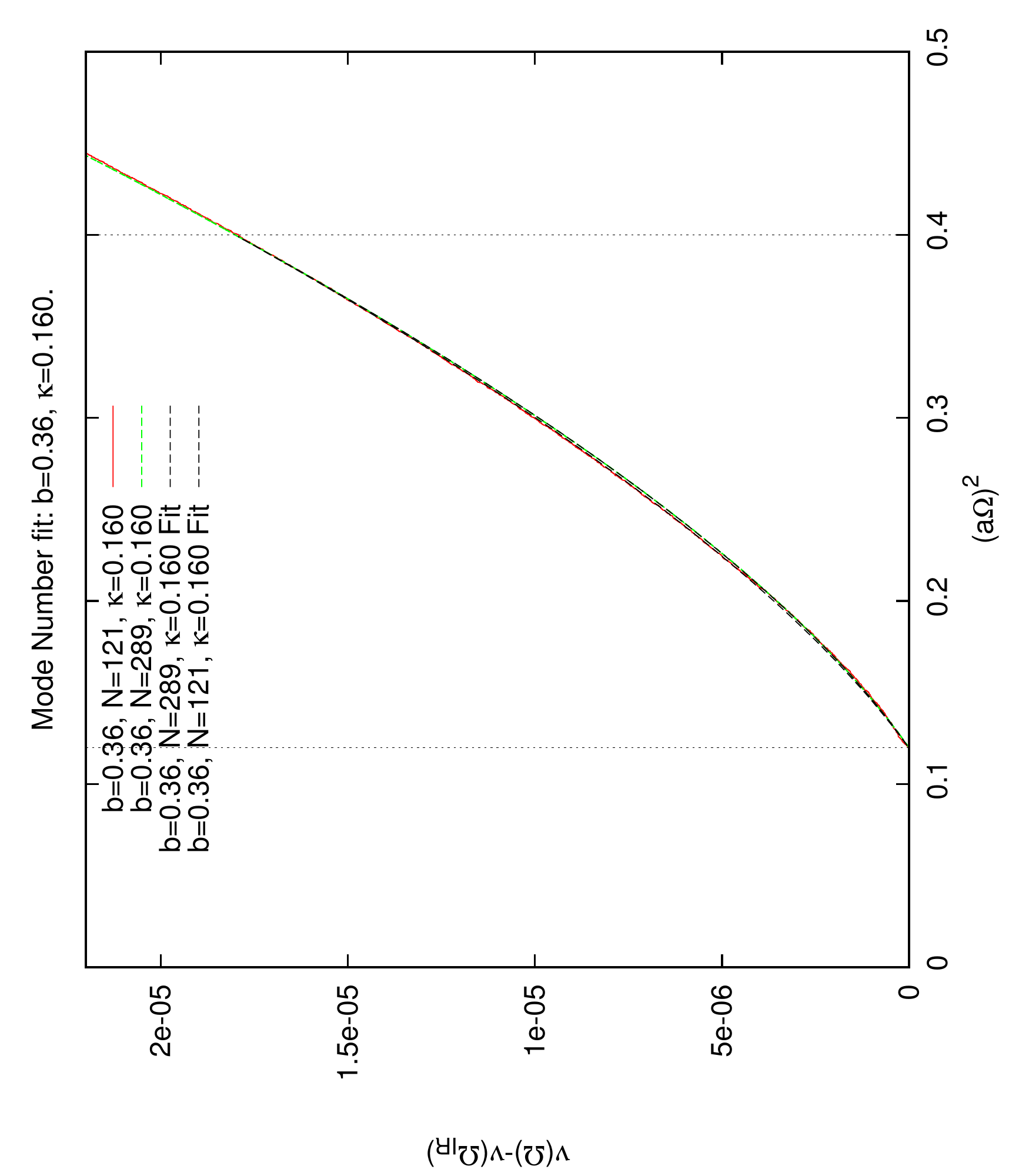}
  \caption{Mode number fit for $b=0.35$ (left) and $b=0.36$ (right), $\kappa=0.160$, $N=121$ and $N=289$ in the range $0.12 < (a\Omega)^2 < 0.40$.}
  \label{fig:fit}
\end{figure}

Fig.~\ref{fig:fit} shows a fit to Eq.~\ref{eq:fitIII}, with $am$ fixed to the values determined in Tab.~\protect\ref{tab:am}, in the range $0.12 < (a\Omega)^2 < 0.40$. The lower bound is chosen to exclude the region of low eigenvalues which suffer from finite volume effects, and the upper bound is chosen to use all the available eigenvalues for $N=289$. The resulting values for $\gamma_*$ are shown in Tab.~\ref{tab:gamma}.

\begin{table}
\begin{center}
\begin{tabular}{c|c|c|c}
b & $\kappa$ & $N$ & $\gamma_*$\\
\hline
0.35 & 0.160 & 289 & 0.262(2)\\
& & 121 & 0.274(5)\\
\hline
0.36 & 0.160 & 289 & 0.222(2)\\
& & 121 & 0.244(10)
\end{tabular} 
\end{center}
\caption{$\gamma_*$ as determined from a fit to the mode number in the range $0.12 < (a\Omega)^2 < 0.40$. The error is only statistical, and does not include the systematic uncertainty due to the choice of range in which to fit.}
\label{tab:gamma}
\end{table}

\begin{figure}
  \centering
    \includegraphics[angle=270,width=7.0cm]{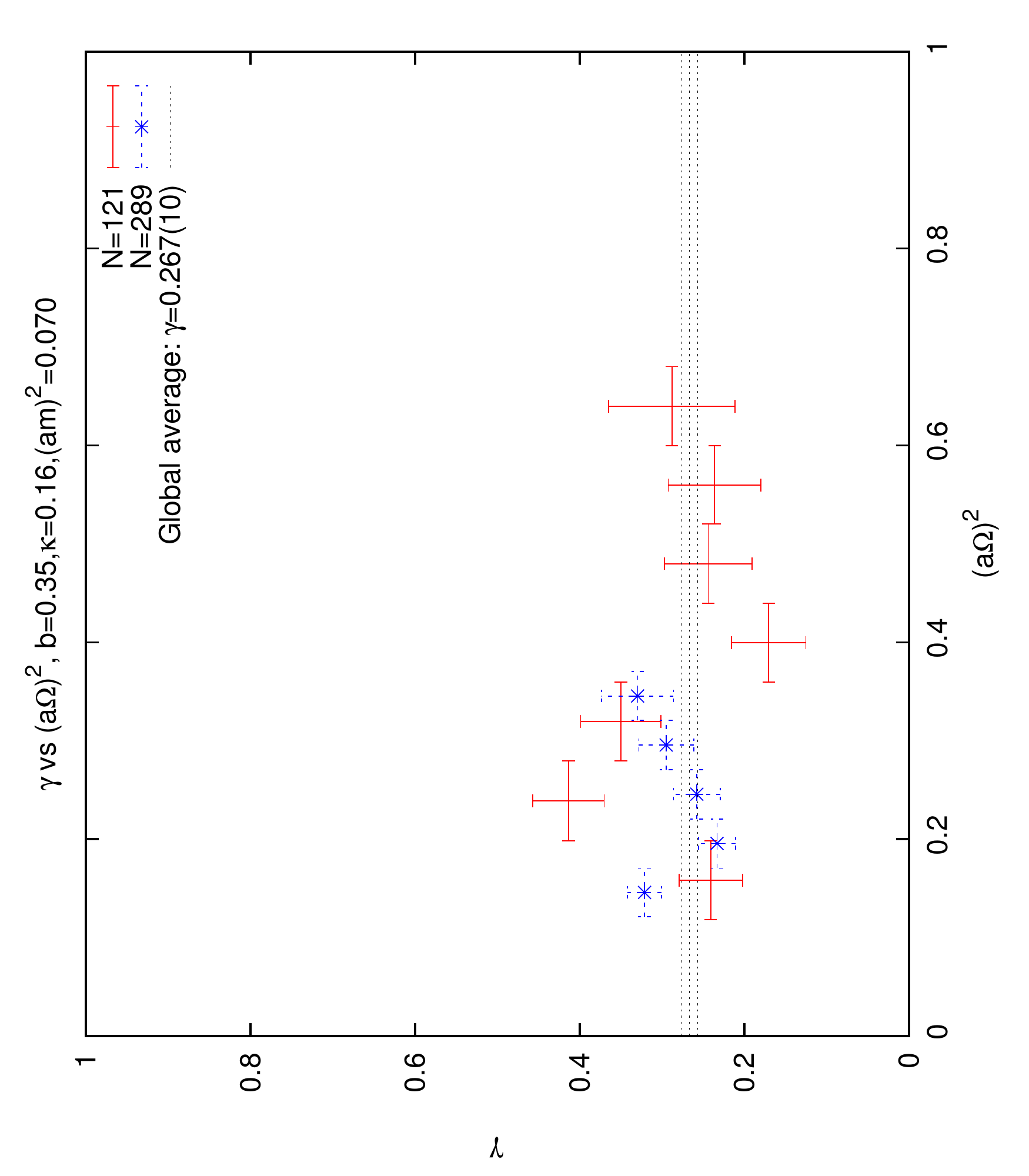}\includegraphics[angle=270,width=7.0cm]{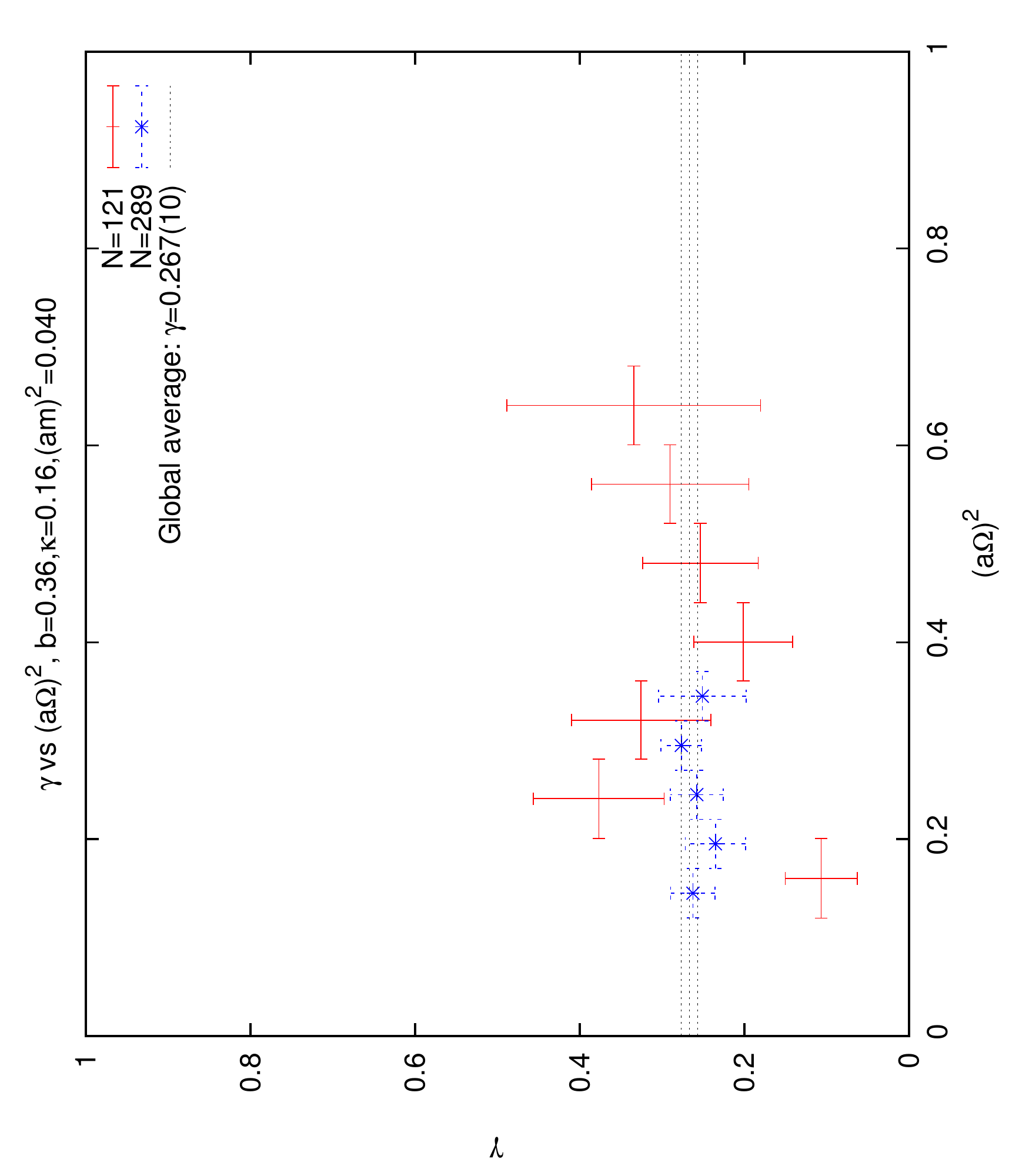}
  \caption{Fitted values for $\gamma_*$ for $b=0.35,0.36$, $\kappa=0.160$, $N=121$ and $N=289$, determined from fits to Eq.~\protect\ref{eq:fitIII}, with $am$ fixed to the values determined in Tab.~\protect\ref{tab:am}. The x error bar shows the fit range, the y error bar shows the statistical error. As there is little apparent dependence on the fit range or $b$, we average over all the $N=289$ data to determine a single value of $\gamma_*=0.267(10)$, shown as the dotted line.}
  \label{fig:fit_2param}
\end{figure}

In order to investigate the dependence of these results on the chosen fit range, Fig.~\ref{fig:fit_2param} shows the value of $\gamma_*$ found for many different fit ranges. In this plot the x error bar shows the fit range, the y error bar shows the statistical error. By taking a much smaller fit range, the statistical errors increase, but for all the fit ranges $\gamma_*$ is broadly consistent with the determination using a wide fit range, and there is no sign of it becoming systematically smaller or larger as we go to larger eigenvalues.

\section{Conclusion}
We simulated the SU(N) gauge theory with two adjoint fermions on a single site lattice using twisted reduction, and from a fit to the mode number of the Dirac operator find $\gamma_*=0.267(10)$. We use two values of the bare coupling, $b=0.35,0.36$, and two values of $N=121,289$. We see agreement between the two bare couplings and values of $N$, and little dependence on the chosen fit range, all of which is consistent with being close to the IRFP.

There is of course no guarantee that we are sufficiently close to the IRFP (i.e. that we have a sufficiently large enough volume and a small enough mass) such that what we measure is actually $\gamma_*$ at the IRFP. It would be interesting to investigate how $\gamma$ changes for a range of values of $b$. We are prevented from going to stronger coupling by the strong coupling phase transition, so investigating larger physical volumes and smaller masses will require larger values of $N$, or possibly using a $2^4$ lattice instead of a single site.

\appendix
\section*{Acknowledgments}
We acknowledge financial support from the MCINN grants FPA2009-08785, FPA2009-09017, FPA2012-31686 and FPA2012-31880, the Comunidad Aut\'onoma de Madrid under the program HEPHACOS S2009/ESP-1473, the European Union under Grant Agreement PITN-GA-2009-238353 (ITN STRONGnet), and the Spanish MINECO's ``Centro de Excelencia Severo Ochoa'' Programme under grant SEV-2012-0249, as well as the use of the SR16000 computer at KEK supported by the Large Scale Simulation Prog.No.12/13-01 (FY2012-13) and the use of the IFT clusters. M. O. is supported by the Japanese MEXT grant No 23540310.

\end{document}